\documentclass[onecolumn,twoside,pra,aps,11pt]{revtex4}
\usepackage{amsmath,mathrsfs,amsbsy,color,graphicx,bm,amsthm,amsfonts,dsfont}
\usepackage{times,wasysym,amssymb}
\usepackage{hyperref}
\usepackage{units}
\usepackage[british]{babel}
\usepackage{epstopdf}

\newcommand{\be}{\begin{equation}}
\newcommand{\ee}{\end{equation}}
\newcommand{\bea}{\begin{eqnarray}}
\newcommand{\eea}{\end{eqnarray}}
\newcommand{\ket}{\rangle}
\newcommand{\bra}{\langle}

\begin{document}
\theoremstyle{remark}
\newtheorem{theorem}{Theorem}
\newtheorem{problem}{Problem}
\newtheorem{proposition}{Proposition}
\newtheorem{example}{Example}

\title{Quantum computation by teleportation and symmetry}

\author{Dong-Sheng Wang}
\affiliation{Stewart Blusson Quantum Matter Institute and Department of Physics and Astronomy, University of British Columbia, Vancouver, Canada}

\begin{abstract}
A preliminary overview of measurement-based quantum computation
in the setting of symmetry and topological phases of quantum matter is given.
The underlying mechanism for universal quantum computation
by teleportation or symmetry are analyzed,
with the emphasis on the relation with tensor-network states
in the presence of various symmetries.
Perspectives are also given for the role of symmetry and phases of quantum matter in
measurement-based quantum computation and fault tolerance.

\end{abstract}


\date{\today}
\maketitle


\section{Introduction}
\label{sec:intr}

Symmetry plays crucial roles in physics.
In quantum many-body physics,
symmetry demonstrates nontrivial interplay with entanglement,
leading to exotic phases of quantum matter~\cite{Wen04,ZCZ+15}.
The understanding of phase transition has also been developed
beyond the spontaneous symmetry breaking paradigm.
Partly motivated by the promising roles in quantum computation (QC)~\cite{NC00,Kit03,NSS+08},
abelian and non-abelian topological (TOP) order~\cite{Wen04},
symmetry protected topological (SPT) order~\cite{CGW11,CGL+12,SPC11},
symmetry-enriched TOP (SET) order~\cite{MR13,LV16},
order with subsystem symmetry~\cite{YDB+18,SSC18,ROW+18,SNB+18},
high-form symmetry~\cite{KS14,GKS+15},
and fracton order~\cite{VHF15,VHF16,HHB17} have been widely investigated in recent years.

QC is a dynamical process from initial input to final output
within a finite amount of time with a certain space cost, hence constrained by symmetry.
The role of symmetry in QC has been studied in various ways,
such as quantum reference frame~\cite{BRS07},
and lately it becomes more and more important in the universal model of QC,
the so-called measurement-based QC (MBQC)~\cite{RB01,Leu04,Nie06,HDE+06,BBD+09,RW12,KWZ12,Wei18}.
Given a quantum resource state, or ‘cluster’,
QC is induced by local projective measurements on it
with outcomes being recorded and feed forward.
The underlying mechanism of MBQC originally was quantum teleportation~\cite{BBC+93,GC99},
which can be viewed as a special case of the much broader framework based on symmetry, emphasized in this work.
The solid relation between MBQC and phases of quantum matter has been demonstrated~\cite{GE07,BM08,DB09,ESB+12,SWP+17}.
It should also been noted that the methods of
measurement and teleportation in MBQC find many other applications,
such as fault-tolerance~\cite{ZLC00}, blind QC~\cite{BFK09}, to name a few.

In this work, we provide a preliminary introduction of the symmetry framework of MBQC.
First a brief review of teleportation
with the focus on symmetry and its relation with tensor-network states
is given in section~\ref{sec:review}.
Then in section~\ref{sec:sym} we discuss two points of view of MBQC,
one is the traditional circuit logic, and the other is the setting of symmetry.
A perspective of roles of various symmetries is given in the conclusion.

\section{Computation by teleportation}
\label{sec:review}

\begin{figure}
  \centering
  \includegraphics[width=\textwidth]{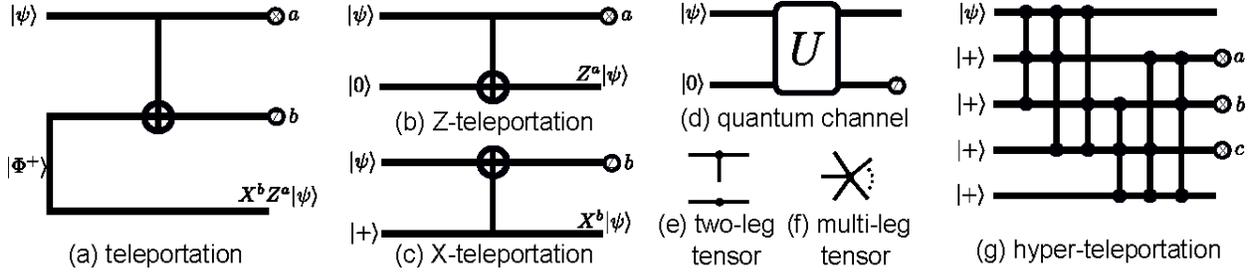}
  \caption{(a) Quantum teleportation of an arbitrary qubit state $|\psi\rangle$.
  The vertical line labels the Bell state $|\Phi^+\rangle$,
  the $\oplus$ with a vertical line labels the CNOT gate,
  the circle with X (Z) in it labels measurement in Pauli $X$ ($Z$) basis,
  and $a$ ($b$) is the binary outcome resulting the byproduct $X^bZ^a$.
  (b) Z-teleportation derived from the swap operation between $|\psi\rangle$ and $|0\rangle$.
  (c) X-teleportation, analog with the Z-teleportation.
  (d) Quantum circuit for a quantum channel from Stinespring's dilation theorem.
  The `ancilla' has initial state $|0\rangle$ and is traced out at the end.
  (e) A two-leg tensor with its physical index shown as a vertical leg (up)
  or absorbed by the dot (down).
  The leg to the left (right) is for the input (output).
  The dot with the physical index $i$ labels the set of tensors $\{A_i\}$.
  (f) A multi-leg tensor.
  (g) Quantum hyper-teleportation with CCZ gates of an arbitrary qubit state $|\psi\rangle$.
  Each vertical line with three dots labels a CCZ gate,
  and a dot labels the nontrivial action on the underlying wire.
  }\label{fig:circuit}
\end{figure}

Here we study the role of teleportation in QC.
Quantum teleportation, or teleportation for brief,
is a unique quantum scheme that does not have classical analog.
The central mechanism of teleportation is that,
given an arbitrary input state $|\psi\ket$ and a Bell state $|B\ket$,
which is a maximally entangled state from the Bell basis
$\{|\Phi^\pm\ket, |\Psi^\pm\ket\}$ for
\be |\Phi^\pm\ket=\frac{1}{\sqrt{2}}(|00\rangle\pm |11\rangle),\;
|\Psi^\pm\ket=\frac{1}{\sqrt{2}}(|01\rangle\pm |10\rangle),\ee
the state $|\psi\ket$ can be transferred to one of the two parts of the Bell state,
by projecting the other part and the input system onto a Bell state.
From Fig.~\ref{fig:circuit}(a),
depending on the measurement outcome $a$, $b$,
the output state is $|\psi\ket$ except correctable Pauli byproduct $X^bZ^a$.

Teleportation has been generalized in many ways,
such as entanglement swapping~\cite{BBC+93,ZZH+93},
group-theoretic generalization~\cite{BDM+00},
and gate teleportation~\cite{GC99}.
To understand the primary mechanism of teleportation,
we employ the idea of the one-bit teleportation~\cite{ZLC00}.
It turns out teleportation is a sequence of two swap operations,
see Fig.~\ref{fig:circuit}(b),(c).
Given that swap is an obvious operation,
teleportation can be intuitively understood in this way.

The bridge between teleportation and QC is the gate teleportation~\cite{GC99}.
The X-bit teleportation and Z-bit teleportation
each can teleport a U(1) group of gates.
By the concatenation of them,
the whole SU(2) (or SU($d$)) group of gates can be realized
due to the appearance of the Hadamard (Fourier) gate.
The crucial application of gate teleportation is fault tolerance~\cite{ZLC00},
as gates can be classified in the Clifford hierarchy (CH)~\cite{GC99}
according to the degree of difficulties for realization.
Gates like the T gate on higher levels of the CH
can be realized by gates on lower levels of the CH
acting on a certain `magic state'~\cite{ZLC00}.
The lowest level of the CH is the Pauli group,
$\mathcal{P}=\bra i\mathds{1},X,Z\ket$ for a single qubit.
The $k$-th level of the CH is a set
\be \mathcal{C}^{(k)}:=\{U| UPU^\dagger \in \mathcal{C}^{(k-1)}, \forall P\in \mathcal{P}_n \} \ee
for $\mathcal{P}_n$ as $n$-qubit Pauli group.
We can see that the byproduct operators in teleportation is from the Pauli group.
Please note we mostly focus on qubit case instead of qudit case for simplicity.

It has been established that gate teleportation can be used for
universal QC~\cite{GC99,RB01,Leu04,AL04,CLN05},
which is usually known as `teleportation-based' or `measurement-based' QC.
The universal set of gates is usually chosen to be $\{$CZ, G$\}$,
where G is the whole group of SU(2),
and CZ is the controlled-Z gate.
If fault-tolerance is required, the gate set can be replaced by $\{$CZ, H, T$\}$,
for H as Hadmard gate,
and the T gate can be realized by magic-state injection~\cite{ZLC00,BK05}.
At the same time, another universal set of gates $\{$CCZ, H$\}$ has been studied in recent years~\cite{Shor96,Shi02,PR13,TYC17,GGM18},
for CCZ as the controlled-CZ gate equivalent to Toffoli gate up to H gate.
This is also motivated by the so-called hypergraph states~\cite{QWL+13,GCS+14,LUW+15}.
Instead of the CZ gate, it is found that the CCZ gate can also be used
for teleportation, see Fig.~\ref{fig:circuit}(g).
The output for the first and last qubits is a state
\be |\Phi\ket=(\mathds{1}\otimes Z^{a\oplus b\oplus c}) (\mathds{1}\otimes \text{H}) |C\ket\ee
for  $|C\ket=\text{CZ} |\psi\ket |+\ket$ usually appeared in MBQC.
A model for universal QC based on the gate set $\{$CCZ, H$\}$ is proposed lately~\cite{GGM18}.
The `hyper'-teleportation in Fig.~\ref{fig:circuit}(g) can be viewed as an encoded teleportation,
however, it does not allow gate teleportation,
which, fortunately, is not necessary for universal QC with the set $\{$CCZ, H$\}$.

\section{Computation by symmetry}
\label{sec:sym}

Here we study the connection between symmetry and QC.
First, there is a symmetry point of view of teleportation.
In the one-bit teleportation, the system and ancilla form a bipartite system.
If one of them is traced out (or projected out),
there is a quantum channel acting on the other.
As is well known, a quantum channel $\mathcal{E}$
can be represented by a set of Kraus operators $\{A_i\}$~\cite{Kra83,Sti55},
and $\sum_i A_i^\dagger A_i=\mathds{1}$ due to the trace-preserving condition.
In teleportation each measurement outcome $i$ is recorded,
therefore, the channel is `selective' or a quantum instrument~\cite{DL70}.
A compact way to represent this is the tensor-network or matrix-product state~\cite{PVW+07}.
The one-bit teleportation can be represented by a two-leg tensor,
see Fig.~\ref{fig:circuit}(d),(e).
Note we ignore the physical leg in the leg-number counting.

\begin{figure}
  \centering
  \includegraphics[width=.8\textwidth]{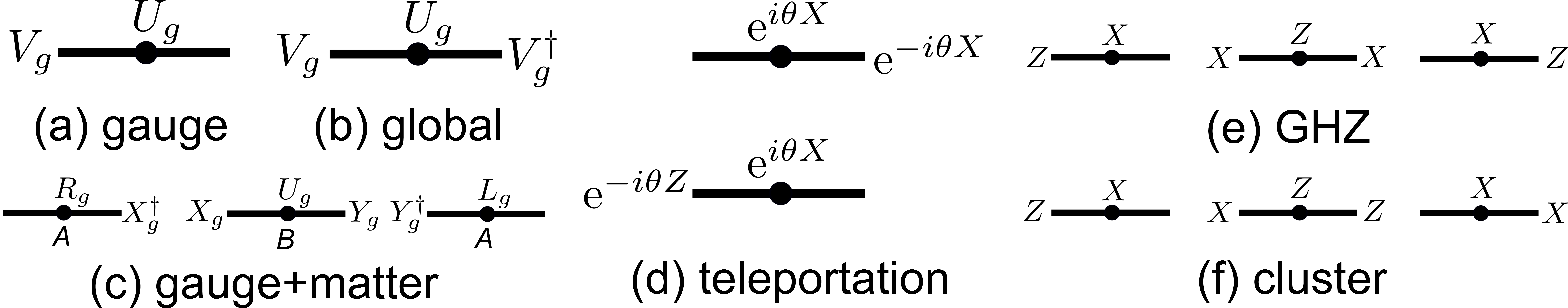}
  \caption{(a) Gauge symmetry with nontrivial action on one leg.
  (b) Global symmetry with the action by conjugation.
  (c) Coupled gauge-matter system.
  (d) Teleportation from U(1) gauge symmetry.
  (e) Symmetries of GHZ state tensor $A_0=\mathds{1}$, $A_1=Z$.
  (e) Symmetries of 1D cluster state tensor $A_0=\text{H}$, $A_1=\text{H}Z$.
  }\label{fig:sym}
\end{figure}

For the one-dimensional case tensors can be connected to form
matrix-product states (MPS)~\cite{PVW+07}.
In recent years MPS with symmetry have been studied
mainly in the setting of symmetry-protected topological (SPT) order~\cite{CGW11,SPC11,CGL+12,DQ13a,KMZ+17}.
The symmetry of a state is global if it acts uniformly on the whole state,
local (gauge) if it acts locally,
or sub-local (intermediate) for the rest cases.
The symmetry of one-bit teleportation is a local symmetry:
it only has nontrivial action on one of the two legs, see Fig.~\ref{fig:sym}(a),(d).
In this picture, the tensor is $A_0=\text{H}$, $A_1=\text{H}Z$
(ignoring the factor $\frac{1}{\sqrt{2}}$),
the gate teleportation can be viewed as a U(1) symmetry.
If the physical leg is measured in a rotated basis by gate $\text{e}^{i\theta X}$,
the output state is acted upon by the nontrivial gate $\text{e}^{i\theta X}$
(or $\text{e}^{i\theta Z}$ on the other leg).

For global symmetry $G$, the action on the legs of tensor $A$ is a conjugation
\be U_g A = V_g A V_g^\dagger ,  \; \forall g\in G, \ee
and $U_g$ is usually a linear unitary representation of the group $G$,
and $V_g$ is a projective unitary representation.
Now if the physical legs of the tensors on different sites
are measured in different bases,
the virtual legs will be acted upon by a sequence of different $V_g$,
which, eventually, form the whole group $G$.
This can be viewed as a scheme of `computation by symmetry'~\cite{WSR17}.
For instance, the spin-1 1D AKLT state~\cite{AKLT87} has global SO(3) symmetry,
which supports universal QC on a single qubit.
Furthermore, there are also states with both global and local symmetries,
such as coupled gauge-matter system~\cite{KMZ+17},
the symmetry action on gauge tensor ($A$) and matter tensor ($B$)
has to be consistent, see Fig.~\ref{fig:sym}(c).
The three-body term $R_gU_gL_g$ plays the role of stabilizers of the state.
The tensor $A_0=\mathds{1}$, $A_1=Z$ defines a GHZ state,
and its symmetry is shown in Fig.~\ref{fig:sym}(e).
With a slight but nontrivial modification,
the 1D cluster state is obtained, Fig.~\ref{fig:sym}(f).
The difference between GHZ state and cluster state is nontrivial,
which will be discussed in the next section.

\section{Multi-angle views of MBQC }
\label{sec:gaklt}

\subsection{Logic point of view}


Here we introduce MQBC from the point view of circuit logic.
In classical computation
Boolean circuits based on Boolean algebra computes
Boolean function $f: \{0,1\}^n \rightarrow \{0,1\}$.
The size of a circuit is usually defined as the number of logic gates, such as NOT, AND etc,
and the depth of a circuit is the maximal number of gates involved
in a path from an input to an output.
As every Boolean function can be written as a Boolean formula in a normal form~\cite{KSV02},
a Boolean circuit can be represented as a tree,
which, in particular, requires the copy operation.

Classical bits can be copied or cloned, however, this is not the case for qubits~\cite{WZ82,Die82}.
It turns out the proper mechanism for copying quantum information is
by entanglement and teleportation.
Given a quantum circuit with $n$-qubits and a certain number of quantum gates,
the circuit depth can be reduced to one by using one-bit teleportation:
suppose gates sequence $ U_n\cdots U_2U_1$ acts on a qubit,
then before gate $U_2$ we attach a qubit ancilla and teleport the qubit to the ancilla,
and the ancilla will play the role of the original qubit.
It is easy to see that the time cost of executing a sequential gates
will be converted to space cost due to the usage of ancilla
with at most a polynomial overhead.
If CZ is the only entangling gates being used and nontrivial qubit gates
are absorbed in measurement bases,
the scheme above is just the same as MBQC,
which prepares a graph state first,
and computes by local projective measurements.

The universality of MQBC on a general resource state
is usually shown by the simulation of a universal set of gates.
Many universal resource states are graph-state like~\cite{HDE+06,VdM+06,VdD+07}.
To simulate a quantum circuit, the `coupled-wire' scheme is usually employed
that each qubit is assigned to a `wire' in a graph state,
and each entangling gate is assigned to a nontrivial junction (by CZ gates) between wires.
Qubit gates can be simulated by gate teleportation along each wire.
The low-depth feature of MBQC has practical benefits:
if each physical qubit has short coherence time,
then MBQC can be employed with the resource state being created as the QC goes.
This has been significantly applied to QC in linear optics by photons~\cite{BR05,OB07}.

In different physical settings, mainly many-body systems,
their ground states as tensor-network states also prove to be natural resource states for MBQC~\cite{GE07}.
By treating a spatial dimension as the simulated time,
and encoding quantum information in the virtual space,
local measurement will enact a gate from the local tensor.
For instance, given a MPS with local tensors $A_{i}$,
a sequence of measurements in bases specified by angles $\phi_i$
will execute gates $\prod_i A(\phi_i)$
(the byproduct issue is ignored for simplicity).
Each tensor can be viewed as a generalization of teleportation, see Fig.~\ref{fig:circuit}.
The state of ancilla in virtual space is transferred from site to site,
and each local particle in the system only interacts with the ancilla once.
With the quantum channel form of a tensor from Fig.~\ref{fig:circuit}(d),
it is easy to see that a MPS can be prepared by a quantum circuit
for an ancilla interacting sequentially with the particles in the system~\cite{SSV+05,WSR17}.
If we further exchange the state of ancilla and particle after each interaction,
and measure the ancilla, then the circuit is an obvious analog of a sequence of gate teleportations.
This generalizes to tensor-network states by cutting MPS wires out and junctions between them for entangling gates.

\subsection{Symmetry point of view}

There has been significant progress by putting MBQC in the setting of many-body physics.
The resource state is usually assumed to be the ground state of a system,
and QC is executed by local projective measurement on each site.
The universality of a state can be shown either directly,
or by reducing it to a graph state~\cite{CDJ+10}.
From the complementary point of view,
it is largely unknown whether a given many-body system is universal or not for MBQC.
Below we study the cases when symmetry is present.
In particular, we do not consider antiunitary symmetry or point group in this work.

It is well established that ground states of many-body system
can be expressed as matrix-product or tensor-network states.
The types of symmetry (global, local, or subsystem),
how the symmetry is represented,
and whether the symmetry is broken or preserved
all provide nontrivial constraints of the tensors therein.
For instance, the GHZ state defined by the tensor in Fig.~\ref{fig:sym}(e) is a ground state
of the 1D Ising model $H=-\sum_i X_i X_{i+1}$.
The model has a global $Z_2$ symmetry $\bigotimes_i Z_i$
which is spontaneously broken.
As has been discussed above,
the GHZ state can be used to execute a U(1) group of gates,
which is not universal on a single qubit.
However, if a `gauging' mechanism~\cite{YDB+18,SSC18} is employed,
the cluster model $H=-\sum_i X_{i-1} Z_i X_{i+1}$ is obtained.
It has a unique ground state as the cluster state defined by the tensor in Fig.~\ref{fig:sym}(f).
The cluster state preserves a global $Z_2\times Z_2$ symmetry,
hence has SPT order,
with one from the global symmetry of the Ising model,
and the other from the gauge subsystem.
When measured in the symmetry basis,
the cluster state can realize the Hadamard gate H:
it is a cellular automata of H~\cite{SVW08,ROW+18,SNB+18}.
To realize the full set of gates in SU(2),
the U(1) symmetry of teleportation meets the demand
by measurements in rotated bases, see Fig.~\ref{fig:sym}(d).

The valence-bond solids is a class of model with SPT order
that has been studied for MBQC.
For instance, the 1D spin-1 AKLT state with tensor $A=\{X,Y,Z\}$
is universal for a qubit QC.
Contrary to teleportation and cluster state,
the execution of gates is not deterministic but heralded~\cite{WSR17}
due to the propagation of Pauli byproduct.
The computation on spin-1 AKLT state can be extended in many ways:
to higher spins or spin ladders~\cite{DR96},
to higher-dimensional lattices~\cite{WAR11,Miy11},
and to higher symmetries, including SU($N$), Sp($N$), and SO($N$) groups~\cite{WSR17} .
The computation power follows from the symmetry directly.

From the example above, it is straightforward to conceive that
SPT order is more powerful than spontaneously symmetry-breaking (SSB) order.
The computational power of 1D SPT order is explored recently~\cite{SWP+17}.
Given a 1D system with SPT order by a global symmetry $G$,
it turns out as long as there is a subgroup $Z_d\times Z_d$ of $G$,
for a certain value of $d$,
it can be used for MBQC on a qudit.
For concreteness, we discuss the examples of Haldane phase for spin-1 and spin-$\frac{1}{2}$ cases.
For spin-1 model with global SO(3) symmetry,
such as the bilinear-biquadratic model,
the ground state is AKLT-like in the gapped phase~\cite{Hal83}.
The tensor is \be A_i=B_i\otimes P_i \label{eq:abp}\ee for $i=0,1,2$
and $P_i$ as the three Pauli matrices $X,Y,Z$,
and $B_i$ can have arbitrary size but is slightly constrained by SO(3).
The reason for the presence of $B_i$ is that
the virtual space is from a direct sum of all irreps of SO(3) with half-integer spins,
and furthermore, the Pauli part $P_i$ in the tensor is common for all of them.
For a spin-$\frac{1}{2}$ model that contains the 1D cluster model as a point,
the global $Z_2\times Z_2$ symmetry also leads to tensor (\ref{eq:abp})
but with $P_i$ also containing identity,
and $B_i$ only with free parameters.
Now for QC, instead of teleportation,
the method is based on symmetry and is algebraic:
the tensor $P_i$ is injective~\cite{PVW+07}, i.e., it spans a logical virtual space,
and measurements in slightly-tilted bases from the symmetry basis
can realize the algebra of $P_i$ and the group SU(2),
which is universal for a single qubit.

\begin{figure}
  \centering
  \includegraphics[width=.45\textwidth]{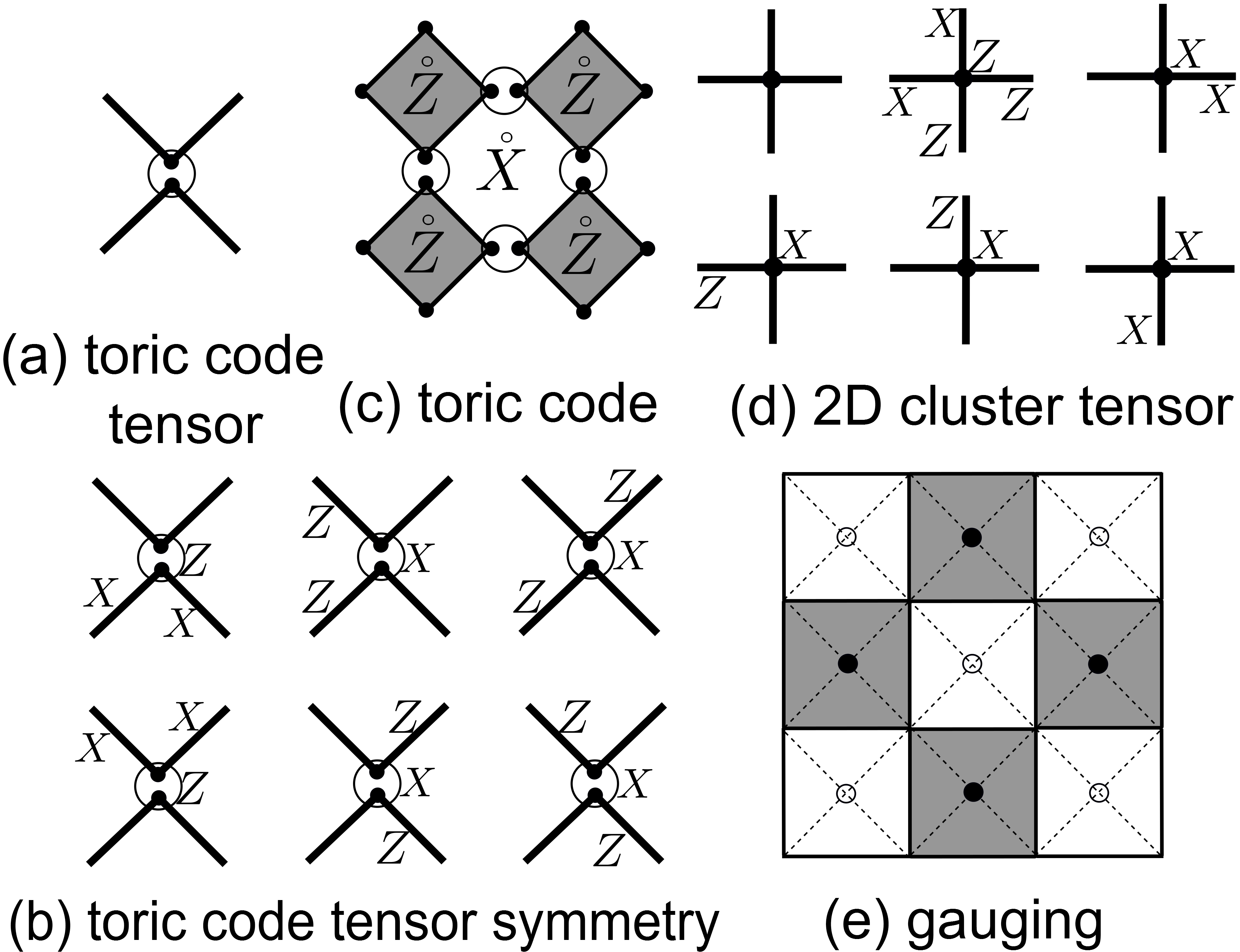}
  \caption{(a) Tensor of the toric code, which is $Z^s\otimes Z^s$ for physical index $s=0,1$.
  The circle represents the physical leg, and each tensor factorizes (two dots).
  (b) Symmetry of the toric code tensor.
  (c) Tensor-network of the toric code.
  The horizontal tensor is the same as the vertical tensor with 90$^\circ$ rotation.
  The $\mathring{X}$, $\mathring{Z}$  stabilizers are shown.
  (d) Symmetry of the 2D cluster state tensor.
  (e) Gauging from toric code to cluster state.
  Gauge qubits (empty dots) are added to the center of each X-cell,
  and then (filled dots) added to the center of each Z-cell.
  The dashed line is the cluster state lattice with a qubit at each vertex.
  }\label{fig:2D}
\end{figure}

With the coupled-wire scheme, the study of 1D SPT order
can be easily extended to higher-dimensional cases with SPT order~\cite{CGW11,SPC11,CGL+12}.
This applies to valence-bond solids and graph states, for instance.
Furthermore, the higher-dimensionality also permits other types of
topological orders.
One notable example is the toric code~\cite{Kit03,Wen03},
which is a $Z_2$ gauge model with purely TOP order.
The topological degeneracy does not depend on system size
except boundary conditions.
It has been argued that toric code is not universal for MBQC
since the computation on it can be efficiently simulated by a
classical computer~\cite{BR07}.
This can also be seen from its tensor structure (Fig.~\ref{fig:2D}(a))
and symmetry (Fig.~\ref{fig:2D}(b)).
Similar with the GHZ state tensor,
there is no Hadamard gate H, hence the group SU(2) on a qubit
cannot be realized.
We denote the four-body X-type (Z-type) stabilizer as $\mathring{X}$ ($\mathring{Z}$)
and the underlying square as X-cell (Z-cell), see Fig.~\ref{fig:2D}(c).
It turns out, analog with the 1D case,
there is also a scheme that plays the role of gauging
leading to a 2D cluster state on square lattice,
which can be viewed as the inverse of the reduction procedure
from 2D cluster state to toric code by measurement.
The tensor and its symmetry for the 2D cluster state is shown in Fig.~\ref{fig:2D}(d).
To show the gauging procedure, as depicted in Fig.~\ref{fig:2D}(e),
we rotate the toric code lattice by $45^\circ$.
The position of each cell is labelled by $n,m$ for its center along the horizontal
and vertical directions.
First, a gauge qubit is added to the center of each X-cell,
and each stabilizer $\mathring{X}_{nm}$ is modified to $\mathring{X}_{nm} Z_{nm}$.
New gauge terms $X_{n,m}ZX_{n\pm1,m\pm1}$ are required
for minimal flux coupling~\cite{Kog79,LG12},
with $Z$ acting on the corner between each two neighboring X-cells.
The stabilizer $\mathring{Z}$ is equal to the product of the four new gauge terms around the Z-cell.
Second, to make the lattice translational invariant,
a gauge qubit is further added to the center of each Z-cell,
and the gauge term $X_{n,m}ZX_{n+1,m+1}$ is modified to
$X_{n,m+1}X_{n+1,m}X_{n,m}ZX_{n+1,m+1}$ and similar for others,
and gauge term $\mathring{X}Z$ is added to each Z-cell.
The resulting cluster state preserves a $Z_2$ subsystem symmetry $\bigotimes_i Z_i$ along
either the horizontal or vertical direction~\cite{EBD12,YDB+18,SSC18,ROW+18},
which is a Wilson loop operator of the toric code,
i.e., a 1-form symmetry~\cite{KS14,GKS+15,Yos16} that is spontaneously broken.
At the same time, the Wilson loop operator $\bigotimes_i X_i$
does not map to a symmetry on the cluster state,
as the excitation by $X_i$ on any site $i$ is static.
The `electric' charge by $X$ is a monopole, i.e.,
it appears solely,
while the `magnetic' charge by $Z$ is a quartet,
and two of them can be shifted away freely from the other two.

In general, the world of two-dimensional systems
with TOP, SET, SPT, and other orders
is largely unexplored for MBQC.
The class of valence-bond solid states with weak SPT order
has been thoroughly investigated~\cite{WAR11,Miy11,Wei18} in recent years.
The family of hypergraph states with strong SPT order
has also been shown to be good resource states~\cite{MM15b,GGM18}.
The quantum dimer models~\cite{RK88},
a seminal model with SET order, however,
have not been studied for the purpose of MBQC.
Criteria and schemes for universal MBQC in these cases need more investigations.

\section{Conclusion}
\label{sec:conc}

In this work we have reviewed the recent research of relating MBQC to symmetry briefly.
This line of research turns out to be fruitful, with intriguing open issues ahead.
The central open problem is that it is unknown what symmetry,
in the setting of many-body systems,
is sufficient (and/or necessary) for universal MBQC.
For instance, there is an apparent mismatch of the treatments of 2D cluster state:
one is based on subsystem symmetry~\cite{YDB+18,ROW+18},
and the other from global symmetry and SPT order~\cite{MM15b}.
Criteria for universal MBQC have been studied from the point of view of entanglement~\cite{VdD+07,GFE09,CDJ+10,VdN13},
and how this study can be improved by (or related to) the symmetry point of view
is an interesting task.
For many-body states with nontrivial entanglement,
the algebraic approach~\cite{SWP+17} seems to be promising:
if the underlying tensors can span a space,
then the space is protected by symmetry and can be used to encode quantum information.

Furthermore, we largely omitted the topic of fault-tolerance in this work,
which yet is an important direction for MBQC.
This has been pioneered in the setting of topological QC with toric code and others~\cite{RHG06,BM06,BM07,BFN08}.
Encoding information with quantum phases of matter,
either by ground states, edge states, or excitations
is supposed to inherit a certain robustness due to the universal features of a phase.
This could provide physical error-resilience for fault-tolerant execution of a discrete universal set of gates.
The efforts of relating MBQC to symmetry so far only focused on universal gate set simulation,
which does not enjoy protection by symmetry or topology.
Therefore, there is a promising research line of bringing MBQC and topological QC together
that can inherit the merits from both sides.
For instance, Majorana-based QC by measurements has been pursued lately~\cite{VF16,AHM+16},
which might turn to be a leading candidate for quantum computers in the near future.

\section{Acknowledgement}

This work is supported by NSERC.
The author acknowledges Z.-C. Gu, Q.-R. Wang,
H. Nautrup, D. Stephen, B. Yoshida, T.-C. Wei, and R. Raussendorf for discussions.

\bibliography{ext}{}
\bibliographystyle{unsrt}
\end{document}